\documentclass[11pt]{article}
\usepackage[normalem]{ulem}
\pdfoutput=1 

\usepackage{jheppub} 
\usepackage{url}
\usepackage{caption}
\usepackage{subcaption}

\usepackage[latin9]{inputenc}
\usepackage{float}
\usepackage{amsmath}
\usepackage{amssymb}
\usepackage{graphicx}
\usepackage{esint}
\usepackage{hyperref}
\usepackage{comment}
\usepackage{color}
\usepackage{microtype}
\usepackage{cleveref}
\usepackage{breakurl}
\usepackage{bbm}
\newcommand{\be}{\begin{equation}}
\newcommand{\ee}{\end{equation}}
\newcommand{\ben}{\begin{displaymath}}
\newcommand{\een}{\end{displaymath}}
\newcommand{\bea}{\begin{eqnarray}}
\newcommand{\eea}{\end{eqnarray}}
\def\K{K{\"a}hler}
   \newcommand{\rf}[1]{(\ref{#1})}

\def\be{\begin{equation}}
\def\ee{\end{equation}}
\def\bea{\begin{eqnarray}}
\def\eea{\end{eqnarray}}
\def\ba{\begin{array}}
\def\ea{\end{array}}
\def\bit{\begin{itemize}}
\def\eit{\end{itemize}}

\newcommand{\N}{\mathcal{N}}
\newcommand{\cN}{\mathcal{N}}

 \makeatletter

\allowdisplaybreaks

\makeatother

\makeatletter
\DeclareRobustCommand{\rcite}[1]{%
  \rcite@aux#1,\@nil{#1}%
}
\def\rcite@aux#1,#2\@nil#3{%
  \if\relax#2\relax
    Ref.~\cite{#3}%
  \else
    Refs.~\cite{#3}%
  \fi
}
\makeatother

\hypersetup{
    colorlinks = true,
    citecolor = {blue},
    linkcolor = {blue},
    urlcolor = {blue},
}

 \title{\rm { \LARGE \bf       Sequestered  Inflation  }}

\author[a]{Renata Kallosh,}
\author[a]{Andrei Linde,}
\author[b]{Timm Wrase,}
\author[c]{and Yusuke Yamada}

\affiliation[a]{Stanford Institute for Theoretical Physics and Department of Physics,\\ Stanford University, Stanford, CA 94305, USA}
\affiliation[b]{Department of Physics, Lehigh University, 16 Memorial Drive East, Bethlehem, PA 18018, USA
}
\affiliation[c]{Research Center for the Early Universe (RESCEU), Graduate School of Science,\\ The University of Tokyo, Hongo 7-3-1
Bunkyo-ku, Tokyo 113-0033, Japan}
\emailAdd{kallosh@stanford.edu}
\emailAdd{alinde@stanford.edu}
\emailAdd{timm.wrase@lehigh.edu}
\emailAdd{yamada@resceu.s.u-tokyo.ac.jp}
\notoc
\preprint{RESCEU-16/21}

\abstract{ We construct supergravity models  allowing  to sequester the phenomenology of inflation from the Planckian energy scale physics. The procedure consists of two steps: At Step~I   we study  supergravity models, which might be associated with string theory or M-theory, and have supersymmetric Minkowski vacua with flat directions. 
  At Step~II we uplift these flat directions to inflationary plateau potentials. We find certain conditions which ensure that the superheavy fields involved in the stabilization of the Minkowski vacua at Step I are completely decoupled from the inflationary phenomenology.  }

\begin{document}

\maketitle

   \newpage

\tableofcontents{}

 \parskip 6pt

\section{Introduction}

Cosmological $\alpha$-attractors \cite{Kallosh:2013hoa,Ferrara:2013rsa,Kallosh:2013yoa,Galante:2014ifa} are among the very few inflationary models favored by the Planck2018 data  \cite{Planck:2018jri,Kallosh:2019hzo}. Simplest versions of these models can match all presently available CMB-based observational data by  tuning of a single parameter. These models can be implemented in the context of $\N = 1$ supergravity, and can be generalized to describe any magnitude of supersymmetry breaking and any value of the cosmological constant \cite{Kallosh:2017wnt,McDonough:2016der}. 

The main advantage of this class of models is the stability of the cosmological predictions with respect to even very significant modifications of the inflationary potential; inflationary predictions are mainly determined by geometric properties of the moduli space. The required geometric properties are the same as in many phenomenological $\N=1$ 4D supergravity descriptions of string theory. This suggests that if one finds a way to suppress the effects related to steep string theory potentials hindering the development of inflation, it may provide us with string theory (or M-theory) inspired versions of the cosmological $\alpha$-attractors.

Thus, our goal is to construct supergravity models which rely on the geometric properties of the string theory moduli space, while sequestering the low energy scale phenomenology of inflation from the Planckian energy scale physics  associated with string theory or M-theory.  This problem is highly non-trivial, because  supergravity potentials  describing many moduli $T_i$ have a lot of mixing terms. Cosmological data suggest that the last 50-60 e-foldings of inflation, which are responsible for the formation of the observable part of the universe, occurred at an energy density that is less than $2 \times 10^{-9}$ of the Planck density \cite{Linde:2016hbb}. Thus one may expect that only some symmetries may protect inflation from being affected by the Planckian energy scale physics. Various ideas of symmetries protecting the flatness of the inflationary potential were studied for a long time in string theory and supergravity,  see for example \cite{Gaillard:1986,Gaillard:1995az,Ellis:2020lnc} and \cite{Silverstein:2008sg, Baumann:2010ys, Baumann:2010nu, Dong:2010in, McAllister:2014mpa} and references therein.

In this paper we will discuss  some models which have supersymmetric Minkowski vacua with flat directions protected by shift symmetries. We will describe a procedure of uplifting these flat directions which preserves their flatness at large values of the moduli and allows to implement $\alpha$-attractors in this context.
 
This procedure consists of two steps: At Step I  we use 4D supergravity with $K^{(I)}, W^{(I)}$ and some number of chiral multiplets $T_{i}$ and find string inspired superpotentials such that the scalar potentials have supersymmetric Minkowski vacua with flat directions. This requirement is satisfied for any \K\ potential if the superpotential $W(T_{i})$ and  its derivatives $\partial_{j}W(T_{i})$ vanish along some direction. 

At Step II we add some $K^{(II)}, W^{(II)}$ that uplift these flat directions to inflationary plateau potentials. We find certain conditions which ensure that  the superheavy fields, which are involved in the stabilization of the Minkowski vacua at Step I, do not affect inflation. Note that the choice of inflationary potentials here is purely phenomenological, we do not derive them from string theory. However, an important property of $\alpha$-attractors is the stability of their most important predictions with respect to the choice of inflationary potentials. Therefore, we hope  that this approach has some merits.

This method was already used in \cite{Gunaydin:2020ric} for constructing inflationary models in M-theory. However, this was part of a large paper relying on the use of technical tools specific for M-theory. 
Meanwhile the main idea of our method is quite general, and it is not limited to M-theory, string theory, or $\alpha$-attractors.
In this paper, we will explain and generalize this  approach, giving some simple examples  to illustrate the general idea of protecting the low energy scale phenomenology of inflation from the Planckian energy scale physics.

\section{Single field models}
\subsection{Step I}
Consider a model  of $n$ chiral superfields $T_{i}$ with a superpotential $W^{(I)}(T_{i})$ (the superpotential at Step I) and with a \K\ potential given by some real holomorphic function $K^{(I)}(T_{i},\overline T_{i})$. If at some point $T_{i} = t_{i} + {\rm i} a_{i}$ the superpotential $W^{(I)}(T_{i})$ and all of its first derivatives vanish,
\be\label{Mink}
W^{(I)}(T_{i})= 0 \ , \qquad  \partial_{T_j}W^{(I)}(T_{i}) = 0 \ ,
\ee
this state corresponds to a stable  supersymmetric Minkowski vacuum  with  vanishing scalar potential $V^{(I)}(T_{i})=0$. This property does not depend on the choice of the \K\ potential.

We will be interested in superpotentials such that the conditions \rf{Mink} are satisfied not only at a single point, but along some flat directions.  The potential may have several different flat directions with $V(T_{i})=0$ \cite{Gunaydin:2020ric}, but the simplest possibility discussed in all examples given in this paper is that the flat direction appears when all fields $T_{i}$ are  equal to each other, and real,   $T_{i} = t$, $a_{i} = 0$.

In this section we will illustrate our general approach using a theory of a single field $T$ with  superpotential $W^{(I)}(T)$, for several different choices of the \K\ potential $K^{(I)}(T,\overline T)$.  One should keep in mind that in the theory of a single field supersymmetric flat directions are possible only if  $W^{(I)}(T)$ vanishes for all $T$, i.e., if $W^{(I)}(T)=0$.   Nevertheless we will keep $W^{(I)}(T)$ in our equations because most of the  results to be obtained in this section can be easily generalized for  models with many fields $T_{i}$, where flat directions satisfying eqs.~\rf{Mink} may appear for non-trivial superpotentials $W^{(I)}(T_{i})$.

\subsubsection{Models with canonical \K\ potential}

Consider the model with superpotential $W^{(I)} (T)$ with  \K\ potential
\be\label{Kcanflat}
K^{(I)}(T,\overline T) =-{1\over 2}  (T - \overline{   T})^{2} \ .
\ee
Now we introduce a nilpotent field $X$, and modify the \K\ potential
\be\label{K1}
K^{(II)} =K^{(I)}(T,\overline T) + {F_X^2 \over F_X^2 +   V_{\rm infl}(T,\overline T) } X\overline X \ .
\ee
We also modify the superpotential
\be\label{WII}
 W^{(II)}=  W^{(I)}+ W_0+ F_X X    \ .
\ee
In the limit $W_0=0$, $F_X= 0$,  and $X=0$ we are back to the  model discussed at Step I. 
Note that \K~potential  $K^{(I)}(T,\overline T)$ and its first derivatives vanish  along its flat direction  $T = \overline T$
\be \label{K0}
\left. K^{(I)}(T,\overline T)\right|_{{T=\overline T }} = 0, \qquad \left.{\partial K^{(I)}(T,\overline T)\over \partial T}\right|_{T=\overline T  } =\left.{\partial K^{(I)}(T,\overline T)\over \partial \overline T}\right|_{T=\overline T }= 0 \ .
\ee  
In this case one can show that if $W^{(I)}(T)$ vanishes along the same direction as $K^{(I)}(T,\overline T)$ (which is trivially satisfied if $W(T)$ vanishes for all $T$), then the potential of the field $T$  for $T=\overline T = t$ is given by
\be\label{infpot}
V_{\rm total}(T) = \left. F_X^2 -3W_0^2 + V_{\rm infl}(T,\overline T)\right|_{T=\overline T=t } \ .
\ee
Note that in this expression in the limit $W_0=0$, $F_X= 0$, $  V_{\rm infl} =0$,  we are back to the  model discussed at Step I.
Here $F_X$ describes the supersymmetry breaking scale and $W_{0}$ corresponds to the gravitino mass. The term $F_X^2 -3W_0^2$ is the cosmological constant  $\Lambda$ at the minimum (if $ V_{\rm infl}$ is zero there which will be always the case for us). For the simplest potential $V_{\rm infl}(T,\overline T) = m^{2} T \overline T$ this yields the inflaton potential
\be
V_{\rm total}(t) = \Lambda + \frac12 m^{2}\phi^{2} \ ,
\ee
where $\phi=\sqrt{2}t$ is the canonically normalized inflaton. In the context of inflationary cosmology, one may safely use the approximation $\Lambda = 0$.  The inflaton field $\phi$ has mass squared $m_{\phi}^{2} = m^{2}$. The inflaton trajectory $T=\overline T = t\ (=\frac{1}{\sqrt2}\phi)$ is stable with respect to the fluctuations in the orthogonal direction $a$, where $T= \frac{1}{\sqrt2}(\phi+{\rm i}a)$. Fluctuations of the $a$ field have positive mass squared  $m_{a}^{2} = m^{2}(1+\phi^{2}) + 4 W_{0}^{2}$.

Using this method and various potentials $V_{\rm infl}(T,\overline T)$ one can find many inflationary models consistent with the Planck data. However, the results will be very sensitive to the choice of the potential $V_{\rm infl}(T,\overline T) $.  Therefore, in next sections we will describe $\alpha$-attractors where this issue can be alleviated.

\subsubsection{{\boldmath General \K\ potentials}}

The derivation of the results obtained in the previous subsection required that the \K~potential  $K^{(I)}(T,\overline T)$ and its first derivatives vanish for $T = \overline T $. If this condition is not satisfied, one may use one of the two equivalent methods to be discussed now.

\vskip 2mm

{\it  {Method 1}}:  One should make a specific \K\ transformation\footnote{Recall that supergravity Lagrangian is invariant under  \K\,  transformation $K\to K-\log A(T)-\log \overline{A}(\overline T)$ and $W\to We^{A(T)}$, where $A(T)$ is a holomorphic function of chiral fields $T$. This invariance is manifest if one uses the K\"ahler invariant function $G=K+\log|W|^2$.} of the \K\ potential $K^{(I)}(T,\overline T)$ and superpotential $W^{(I)}$ at Step I:
\be
{\cal K^{(I)}}(T,\overline T) = K^{(I)}(T,\overline T)-{\kappa(T)+ \kappa(\overline T)\over 2}\,, \qquad 
{\cal W^{(I)}}(T) = W^I(T)\, e^{ \kappa(T)/2} \ ,
\ee
where
\be
\kappa(T) =  K^{(I)}(T,\overline T)_{|_{\overline T \to T}}\  , \qquad  \kappa(\overline T) =  K^{(I)}(T,\overline T)_{|_{T \to \overline T}} \ .
\label{kappa}\ee 
For example, one may start with the simple \K\ potential  $K^{(I)}=T \overline T$, which is not flat in the  direction $T = \overline T$.  In this case $\kappa(T) = T^{2 }$ and $\kappa(\overline T) =   \overline T^{2 }$. Subtracting ${T^{2 }+ \overline T^{2 }\over 2}$ from $K^{(I)}=T \overline T$  yields the \K\ potential $-{1\over 2}  (T - \overline{   T})^{2}$ \rf{Kcanflat}, which has the flat direction $T = \overline T$. 

The new formulation describes the same theory, with the same flat directions of the superpotential as in the original theory (if there were any), but the  \K\ potential in the new formulation satisfies the required flatness conditions \rf{K0}. Then one can use the same procedure as in the case considered in the previous subsection, with the final result given in eq.~\rf{infpot}. This method was used in  \cite{Gunaydin:2020ric} in application to inflation in M-theory. 

\vskip 2mm

{\it   {Method 2}}:  After making these modifications at Step I and adding the term $W_0+ F_X X$ at Step II to the modified superpotential ${\cal W}(T)$, one can perform a {\it reversed} \K\, transformation:
multiply the superpotential $W_I(T)\, e^{ \kappa(T)/2} + W_0+ F_X X$ by $e^{ -\kappa(T)/2}$, and make the corresponding transformation of ${\cal K^{(I)}}$. {\it This returns the \K\ potential ${\cal K^{(I)}}$ and  the superpotential ${\cal W^{(I)}}(T) $ to their original form $ K^{(I)}$  and $W^{(I)}(T)$}. The only change which appears after this set of procedures is the modification 
\be
W_0+ F_X X \qquad  \Longrightarrow \qquad (W_0+ F_X X)\, e^{- \kappa(T)/2} 
 \ee
 in the expression for $W^{(II)}$ in eq.~\rf{WII}.
 
 {\it The methods 1 and 2 produce equivalent results.} We will use the more compact method 2 in the present paper. Namely at Step II we have
\be
K^{(II)}(T,\overline T)= K^{(I)}(T,\overline T) +   {F_X^2 \over F_X^2 +   V_{\rm infl}(T,\overline T) } X\overline X \ ,
\label{KIIg}\ee
\be
W^{(II)} (T) = W^{(I)} (T) + (W_0+ F_X X)\, e^{- \kappa(T)/2} \ ,
\label{WIIg}\ee
where $\kappa(T)$ is defined in eq. \rf{kappa}. This results in 
\be\label{infpotg}
V_{\rm total}(T) = \Lambda + V_{\rm infl}(T,\overline T)_{|_{T=\overline T = t}} \ ,
\ee
where $\Lambda=F_X^2 -3W_0^2$.

In a theory with many fields $T_{i}$  with the \K\ potential given by a sum of independent \K\ potentials for each field $T_{i}$  one should multiply $W_0+ F_X X$ by a product of terms $e^{- \kappa(T_{i})/2}$ for each of the fields $T_{i}$.

In the next subsection we will apply this method to $\alpha$-attractors.

\subsubsection[$\alpha$-attractors]{\boldmath $\alpha$-attractors}

Here we will consider a \K\ potential which often appears in $\cN=1$  4D supergravity describing string theory phenomenology. Therefore at Step I we take
\be\label{KE}
K^{(I)}(T,\overline T) =- \log  (T + \overline{   T}), \qquad W^{(I)}=0 \ .
\ee
At Step II according to eqs.~\rf{KIIg} and \rf{WIIg} we take
\be
K^{(II)}(T,\overline T)= - \log  (T + \overline{   T})+   {F_X^2 \over F_X^2 +   V_{\rm infl}(T,\overline T) } X\overline X \ ,
\label{KIIattr}\ee
\be
W^{(II)} (T) =  (W_0+ F_X X)\sqrt{2T} \ ,
\label{WIIattr}\ee
since in this model 
$
e^{- \kappa(T)/2}= \sqrt{2T}
$.\footnote{Using a K\"ahler transformation, we find $K^{(II)}=-\frac{1}{2}\log \frac{(T+\overline T)^2}{4T\overline T}+{F_X^2 \over F_X^2 +   V_{\rm infl}(T,\overline T) } X\overline X$, \, $W^{(II)}=W_0+F_XX$. If we take $V_{\rm infl}=0$, this expression manifests the invariance $\phi\to \phi+c$, where $\phi$ is defined in eq.~\eqref{defT}. This shift symmetry was explained in detail in the context of the hyperbolic geometry of $\alpha$-attractors in \cite{Carrasco:2015uma}.}
The difference with the previous case is that the field $T$ will have a non-minimal kinetic term. It is convenient to represent the field $T$ as follows:
\be
T = e^{-\sqrt{2}\phi}(1+{\rm i}\sqrt {2} a)\ .\label{defT}
\ee
The inflaton field $\phi$ is canonical, whereas the field $a$ is also canonical for $a\rightarrow 0$. 
As an example, one may consider a potential 
\be
V_{\rm infl} = m^{2}(1-T)(1-\overline T) \ .
\ee
Along the inflaton direction, the total potential is 
\be\label{Ep}
V_{\rm total}(\phi) =\Lambda +  m^{2}\left(1- e^{-\sqrt{2}\phi}\right)^{2} \ .
\ee
If instead of the \K\ potential \rf{KE} one uses the \K\ potential 
\be
K^{(I)}(T,\overline T) =- 3\alpha \log  (T + \overline{   T})\ ,
\ee 
one should use
\be
W^{(II)} (T) =  (W_0+ F_X X) (2T)^{3\alpha/2} \  
\label{WIIattr2}\ee
to find a family of E-model versions of $\alpha$-attractors with
\be
V_{\rm total}(\phi) =\Lambda + m^{2}\left(1- e^{-\sqrt{2\over 3\alpha}\phi}\right)^{2} =\Lambda + m^{2}\left(1- 2e^{-\sqrt{2\over 3\alpha}\phi}+...\right) \ .
\ee
One can considerably change the potential $V_{\rm infl}(T,\overline T)$, but the large $\phi$ asymptotic behavior of this potential remains the same, up to a change of the parameter $m^{2}$ and a redefinition (shift) of the canonical field $\phi$. This is the reason why these models are called cosmological attractors. The model \rf{Ep} describes $\alpha$-attractor with $\alpha = 1/3$.  In $\cN=1$ supergravity $3\alpha$ is an arbitrary parameter defining the moduli space curvature.  In the case with an underlying maximal supersymmetry or string/M-theory one may have  $3\alpha=7,6,5,4,3,2,1$, see Refs.~\cite{Ferrara:2016fwe,Kallosh:2017ced,Kallosh:2017wnt,Gunaydin:2020ric}.

\begin{figure}[!h]
\vspace{-1mm}
\hspace{-3mm}
\begin{center}
 \includegraphics[scale=0.335]{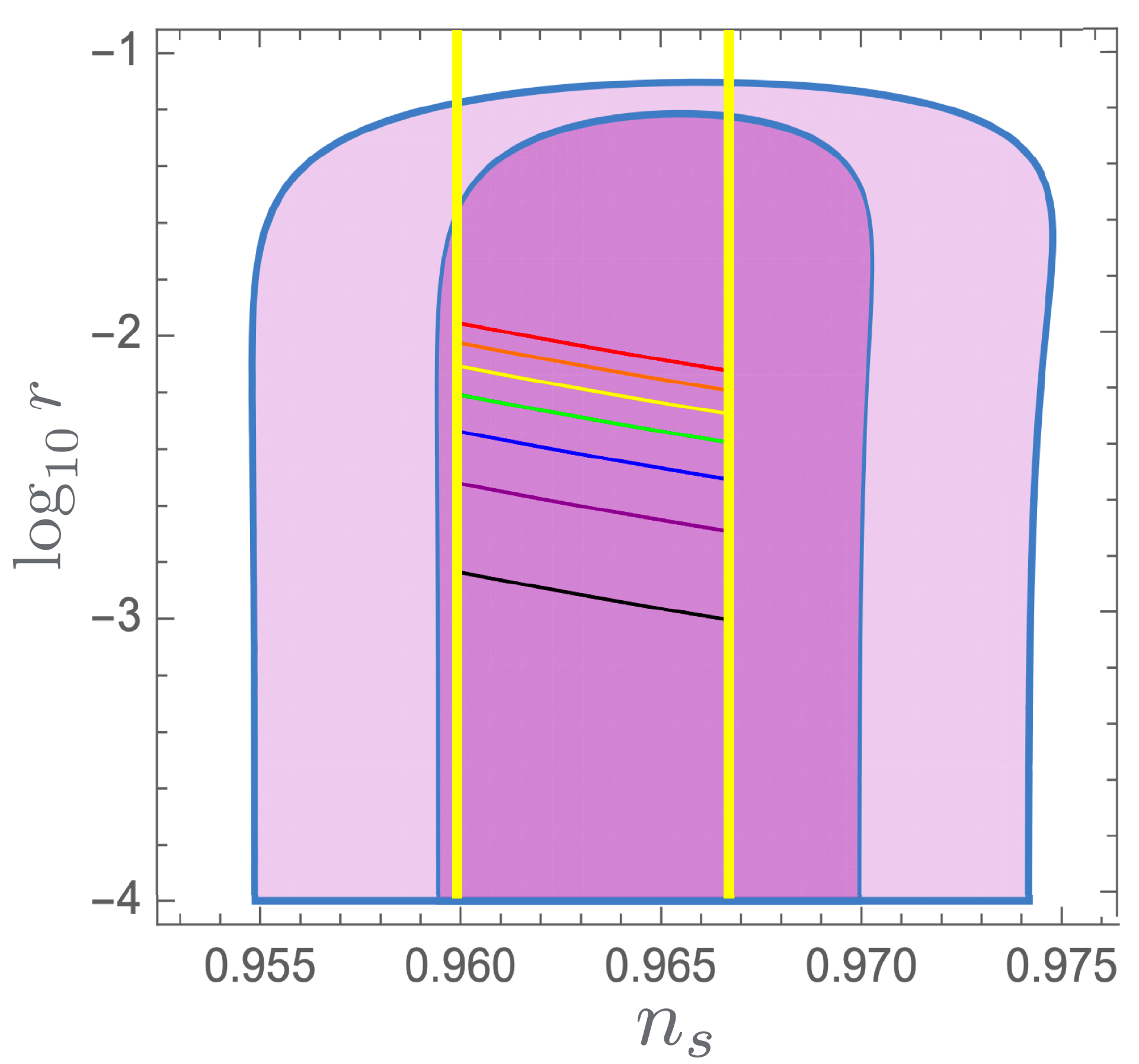}  \hskip 30pt
\includegraphics[scale=0.34]{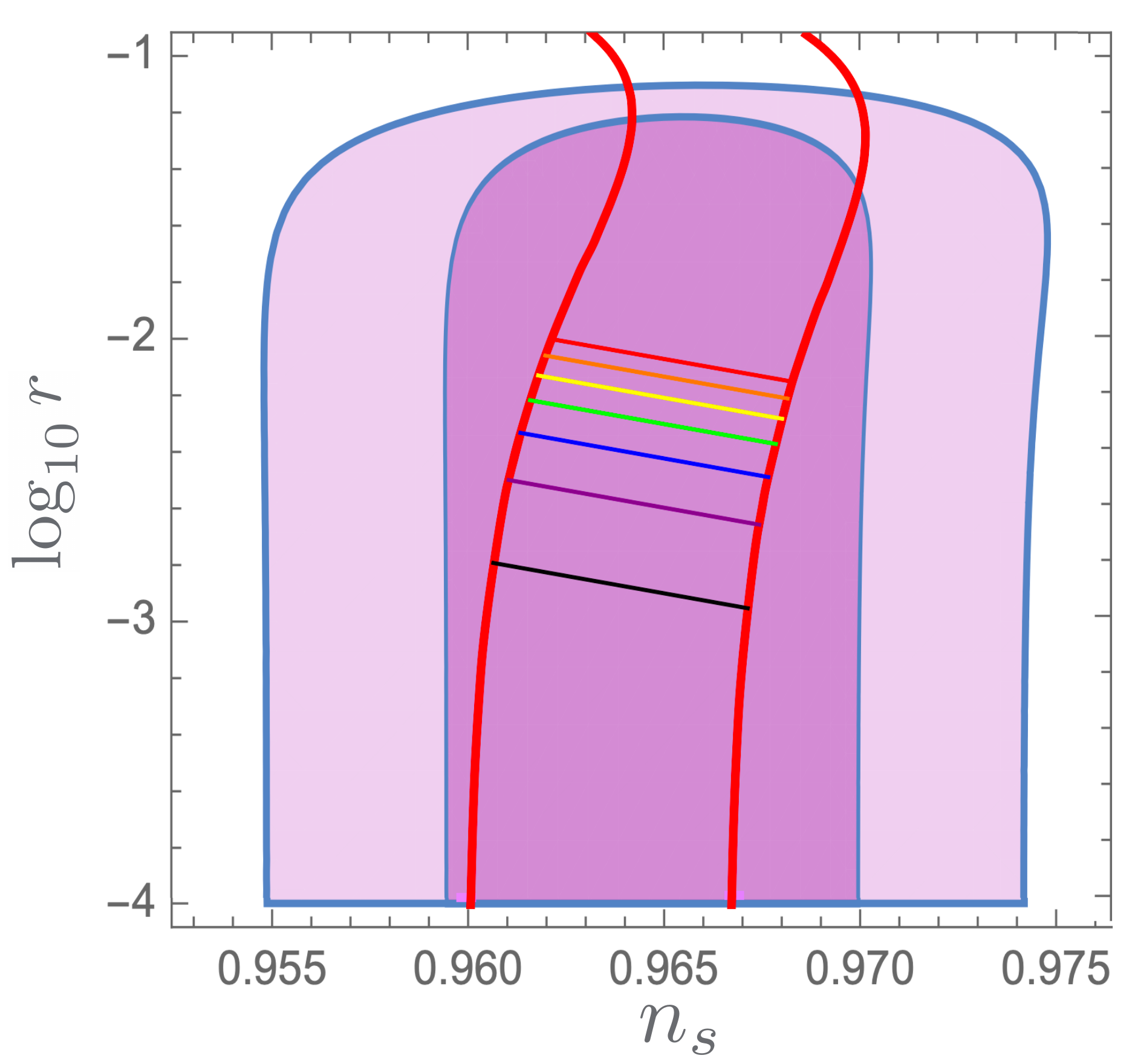}
\end{center}
\vspace{-.12cm}
\caption{\footnotesize
Predictions of the simplest T-models (left) and  E-models (right) for  $50 \leq  N_{e}  \leq 60$ \cite{Kallosh:2019hzo}. Red ellipses show the Planck 2018 results taking into account  the CMB-related data including data from BICEP2/Keck.  String theory motivated versions of these models  have 7 specific predictions \cite{Ferrara:2016fwe,Kallosh:2017ced}: $3\alpha=7$ (red line),  $3\alpha=6$ (orange), $3\alpha=5$ (yellow), $3\alpha=4$ (green), $3\alpha=3$  (blue), $3\alpha=2$ (purple) and     $3\alpha=1$ (black).  All other values of $\alpha$  originate from general $N=1$ supergravity models. }
\label{7disk2}
\end{figure}

Instead of $T$ variables  with \K\ potential  \rf{KIIattr} and superpotential \rf{WIIattr2} describing a half (complex) plane  $T + \overline T > 0$, one can use disk variables $Z=\frac{T-1}{T+1} $ with the resulting  \K\ potential
\be\label{KT}
K^{(II)}(Z,\overline Z) =- 3\alpha \log  (1 - Z\overline{Z})  +   {F_X^2 \over F_X^2 +   V_{\rm infl}(Z,\overline Z) } X\overline X,
\ee
and superpotential
\be \label{WT1}
 W^{(II)}=  (W_0+ F_X X) (1-Z^{2})^{3\alpha/2}   \ ,
\ee
which yields
\be\label{infpot3}
V_{\rm total} (Z) = \Lambda + V_{\rm infl}(Z,\overline Z)_{|_{Z=\overline Z = z}} \ .
\ee
For $V_{\rm infl}(Z,\overline Z)= m^2 Z\bar Z$ (and ignoring $\Lambda$)
this leads to a family of T-models with the potential of the canonical inflaton field  $\phi$
\be\label{Tp}
V(\phi) = m^2 \tanh^{2}{\phi\over \sqrt{6\alpha}} \ ,
\ee
where $z=\tanh{\phi\over \sqrt{6\alpha}}$.  At  $\alpha\ll 1$, E-models \rf{Ep} and T-models \rf{Tp} give very similar predictions for the spectral index $n_{s}$ and tensor to scalar ratio $r$ for a given number of e-foldings $N_{e}$:
 \be
 n_{s}= 1-{2\over N_{e}}, \qquad r = {12 \alpha \over N_{e}^{2}} \ .
  \ee
At larger $\alpha$, the predictions of these two families
 of models slightly differ, see Fig.~\ref{7disk2}.
\vskip -40pt 

\section{Two field model }
\subsection{Step I}
Now we will study a more advanced model, with two fields $T_{1}$ and $T_{2}$, which have a nontrivial superpotential already at Step I: 
\be
K^{(I)}= - {1\over 2} \sum_{i=1}^2 (T_i-\overline T_i)^2\, , \qquad W^{(I)}= M (T_1-T_2)^2 \ .
\ee
There is a supersymmetric Minkowski vacuum at $T_1=T_2$. 
The canonically normalized variables $T_i=\frac{1}{\sqrt{2}}(\phi_i+{\rm i}a_i)$ can be rewritten in terms of canonically normalized mass eigenstates as $\phi=\frac{1}{\sqrt{2}}(\phi_1+ \phi_2)$, $a=\frac{1}{\sqrt{2}}(a_1+a_2)$, $\phi_M=\frac{1}{\sqrt{2}}(\phi_1- \phi_2)$ and $a_M=\frac{1}{\sqrt{2}}(a_1-a_2)$. The mass eigenvalues of $\phi,a$ at $T_1=T_2=0$ are $0$ and the mass eigenvalues of the massive modes $\phi_M,a_M$ at $T_1=T_2=0$ are $m^2=16M^2$. This mass is below the Planck mass $M_{Pl} = 1$ for $M<{1\over 4} $. 

The absence of tensor modes at the level $r \lesssim 0.05$ implies that the inflaton potential $V$ at the last 60 e-foldings of inflation should be smaller than $2 \times 10^{-9} $ \cite{Linde:2016hbb}.  This means that if the field  $T_1+T_2= \phi +{\rm i} a$ deviates from its equilibrium at $T_1+T_2 = 0$ by $O(1)$, its potential can become many orders of magnitude higher than the inflationary potential $V$. Thus it is necessary to  check whether the scenario outlined in the previous section may remain  intact despite the introduction of the large  superpotential $W^{(I)}= M (T_1-T_2)^2$.

\subsection{Step II}
\begin{figure}[!h]
\vspace{-10mm}
\begin{center}
 \includegraphics[scale=0.08]{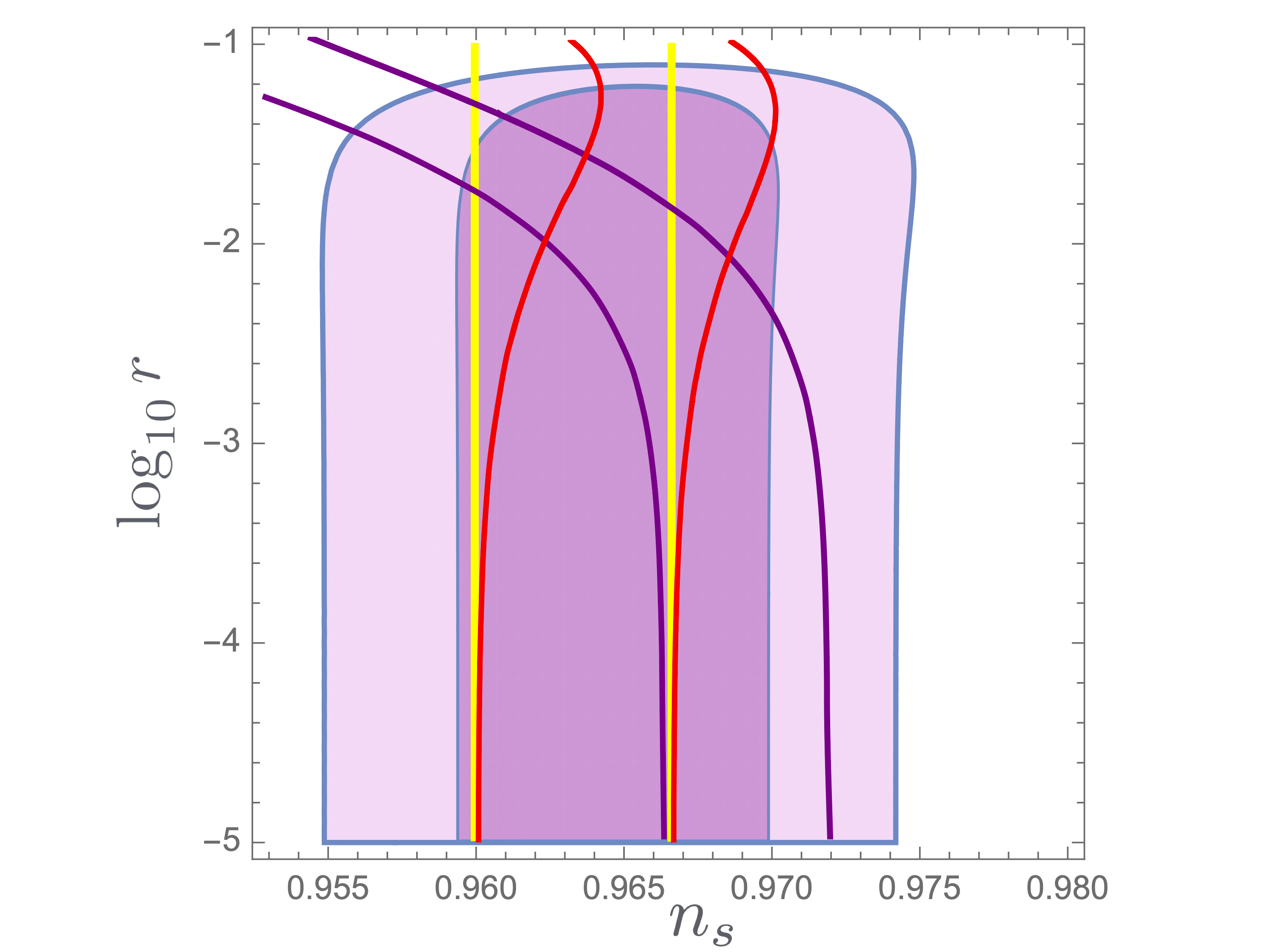} \end{center}
\vspace{-.3cm}
\caption{\footnotesize
A combined plot of predictions of the simplest T-models (bounded by two yellow lines corresponding to $50 \leq  N_{e}  \leq 60$) the simplest E-models (bounded by red lines) and the KKLTI model  (bounded by purple lines) \cite{Kallosh:2019hzo}. }
\label{TEDI}
\end{figure}

Just as in the previous section, we introduce a nilpotent field $X$ and modify the \K\ potential and the superpotential
\bea
K^{(II)}&=& K^{(I)} + {F_X^2 \over F_X^2 + V_{\rm infl}(T_i, \overline T_i) } X\overline X \ ,
\label{KII}\\
 W^{(II)}&=& W^{(I)} + W_0+ F_X X  \ .
\eea
Fortunately, the \K\ potential $K^{(I)}$ vanishes along the flat direction  $T_1=\overline T_1 = t$, $T_2=\overline T_2 = t$ along which the superpotential $W^{(I)}$ also vanishes. As a result, one can show that eq.~\rf{infpot} derived for the model with  $W^{(I)}= 0$ remains valid as well. 
At this time, instead of a simple quadratic scalar potential we will take a more complicated one
\be
V_{\rm infl}(T_i, \overline T_i)=  {g^2 |T_1|^2 |T_2|^2\over \mu^4 + |T_1|^2 |T_2|^2} \ .
\label{Pot1}\ee
At $T_M=0$ and $T_i=\overline T_i = t$ the potential depends only on the canonical inflaton field $T=\frac{1}{\sqrt{2}}\phi\ (=t)$ and (for $\Lambda = 0$) is given by
\be
V_{\rm infl}(\phi) =   g^2 {\phi^4\over 4\mu^4 +\phi^4} \ .
\label{Pot2}
\ee
 This is the plateau potential of the  D-brane inflation model described recently in \cite{Kallosh:2018zsi}. It was  called KKLTI in the Encyclopedia Inflationaris \cite{Martin:2013tda} because of its possible relation to the KKLT mechanism of vacuum stabilization and inflation  in string theory \cite{Kachru:2003aw,Kachru:2003sx}. This model is in good agreement with observational data. Its prediction 
\be 
n_{s} = 1-{5\over 3N_{e}}  
\ee
 is very close to the prediction of the $\alpha$-attractors $n_{s}= 1-{2\over N_{e}}$. This  model and  $\alpha$-attractors almost completely cover  the area in the $(n_{s},r)$ plane favored by Planck2018 \cite{Kallosh:2019hzo}, see Fig. \ref{TEDI}. 

During inflation the mass squared of the heavy fields $\phi_M, a_M$ remain approximately the same as it was at Step I, so it can take any value up to the Planck mass. This means that the fields $\phi_M$ and $a_M$ are strongly stabilized at $\phi_{M}=a_M =0$. The mass squared of the imaginary component $a$ of the inflaton field field $T = {1\over \sqrt 2}(\phi + {\rm i} a)$ is positive.  Thus, the inflaton trajectory $T =   {1\over \sqrt 2} \phi$, $\phi_M=a_M=a = 0$ is stable. Importantly, the potential masses of  the light fields $\phi$ and $a$  and the inflaton potential~\rf{Pot2} do not depend on $M$. This ensures sequestering of the inflaton potential from the  
high energy scale physics related to  the superpotential $W^{(I)}= M (T_1-T_2)^2$.

The development of inflation depends on our choice of potential $V(T_i, \overline T_i)$ in eq.~\rf{Pot1}. Note that one can add to this potential various terms stabilizing the inflaton trajectory  without affecting the inflaton potential $V(\phi)$. For example, any term proportional to $-(T_{i}-\overline T_{i})^2$ vanishes along the inflaton trajectory, but it allows  to increase the mass squared of the axion $a$. We also note that the mass of $T_{M}$ is not much affected by the inflaton dynamics in this model because the K\"ahler potential is independent of the inflaton $\phi$. This might not be the case in more general setups, and we will show an example in the next section.

\section{Three field model }
\subsection{Step I}
The next  model  describes three fields $T_1$, $T_2$ and $T_3$:
\begin{eqnarray}
K^{(I)} &= & -\sum_{i=1}^{3}\ln  (T_{i}+\overline{T}_{i})    \, , \qquad W^{(I)} = {1\over 4} (T_1 - T_2) (T_2 - T_3) \ .
\label{K}\end{eqnarray}
There is a supersymmetric Minkowski flat direction  $T_1=T_2=T_3$, corresponding to a massless field   
\be
T = {1\over 3} (T_1+T_2+ T_3) \ .
\label{flat}\ee
There are also two heavy   fields $T_{M_1}$, $T_{M_2}$, which correspond to some other combinations of the fields  $T_1, T_2,  T_3$. 

The fields $T_{i}$ can be  represented as  $T_i= e^{-\sqrt {2}\, \phi_i}\, (1 +{\rm i} \sqrt 2   a_i)$, where the fields   $\phi_{i}$  and the fields $a_{i}$  are canonical in the small $a_{i}$ limit.   When   $\phi_i ={1\over \sqrt 3} \phi$ and $a_i=0$ the mass squared eigenvalues of the canonically normalized fields $\phi_i$ are:
\be
m^2= 0, \qquad m^2 =\frac98 \, e^{-\sqrt {2 \over 3}\phi}, \qquad m^2=\frac18 \, e^{-\sqrt {2 \over 3}\phi} \ .
\ee
The masses of the heavy fields change depending on $\phi$.  Therefore, we must check  at Step II whether the  heavy fields  remain heavy during inflation and decouple from the inflationary dynamics.

\subsection{Step II}

The \K\, potential and  superpotential are
\bea
K^{(II)} &=& -\sum_{i=1}^{3}\ln  (T_{i}+\overline{T}_{i})  + {F_X^2 \over F_X^2 + V_{\rm infl}(T_i, \overline T_i) } X\overline X \ ,\cr  
W^{(II)} &=& W^{(I)}+ (W_0 + X F_X) \left(\prod _{i=1} ^3 2T_i\right)^{1/2}.
\label{KII3}\eea
Our choice of the potential in equation \rf{KII3} with $T$ defined in equation \rf{flat} is
\be
V_{\rm infl}(T_i, \overline T_i)= \mu^2 \left (1-\frac13 \sum_{i=1}^3T_i\right)\left (1-\frac13 \sum_{i=1}^3\overline{T}_i\right) \quad \to\quad V_{\rm infl}(T, \overline T)=\mu^2(1-T)(1-\overline T).
\label{pot}\ee
In the first approximation, let us follow the supersymmetric trajectory $T_1=T_2=T_3=T$ on which the K\"ahler and superpotential are reduced to
\begin{align}
K= -3\ln  (T+\overline{T})  + {F_X^2 \over F_X^2 + V_{\rm infl}(T , \overline T) } X\overline X, \qquad
W=(2T)^{\frac32} (W_0 + F_X X) .\label{W3eff}
\end{align} 
At  $T_1=T_2=T_3$ and $T_i=\overline T_i$ the potential depends on the field $T=t= e^{-\sqrt {2 \over 3} \phi}$ and is given by
\be
V_{\rm total}(t) = \Lambda + \mu^2 (1-t)^2=  \Lambda + \mu^2 \left(1-e^{-\sqrt {2 \over 3} \phi}\right)^2 \ .
\label{3mod_Pot}\ee
Here with $\Lambda =F_X^2 -3W_0^2>0$, inflation ends at $t=1, \phi=0$ in a de Sitter vacuum. This is the $\alpha=1$ attractor model, whose bosonic part coincides with that of the Starobinsky model. 

\begin{figure}[h!]
\centering
\includegraphics[scale=0.45]{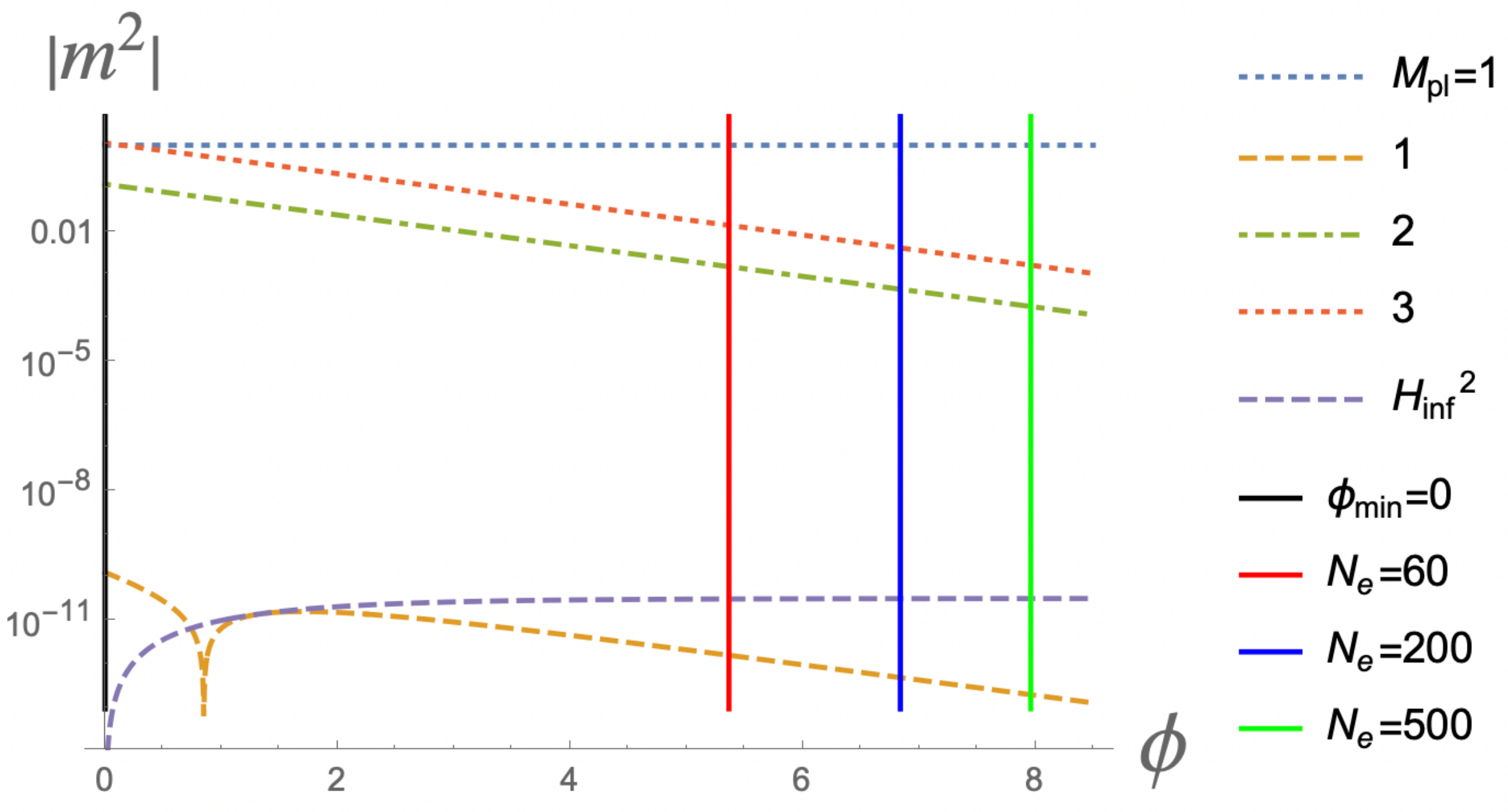}
\caption{\footnotesize  Here we plot, using the logarithmic scale, the absolute value of the masses squared of the heavy fields, the two top lines, not far from the $M_{Pl}=1$ scale. The absolute value of the mass squared of the inflaton field is many order of magnitudes lighter: This is the uplifted flat direction of Step I, the 
inflaton field. The kink at $\phi \approx 0.8$ is due to the fact that  the inflaton mass squared, which was negative during early stages of inflation,  and then vanishes and changes its sign, and we plot logarithm of its absolute value. The actual mass squared of the inflaton field is shown in Fig.~\ref{3mod}.}
\label{3mod}
\end{figure}

In order to make sure that the sequestering mechanism works consistently, we need to check the behaviour of the two heavy fields during inflation.  The masses of them calculated by using eqs. \eqref{KII3} are presented in Fig.~\ref{3mod}, for the choice of parameters $W_0=F_X/\sqrt{3}=10^{-5},\ \mu =10^{-5}$. The axion masses of these heavy superfields are only slightly different from the saxion masses plotted here, this difference is due to the small values of the parameters $W_0=F_X/\sqrt{3}=10^{-5},\ \mu =10^{-5}$.

\begin{figure}[h!]
\vspace{-1mm}
\hspace{-3mm}
\begin{center}
 \includegraphics[scale=0.37]{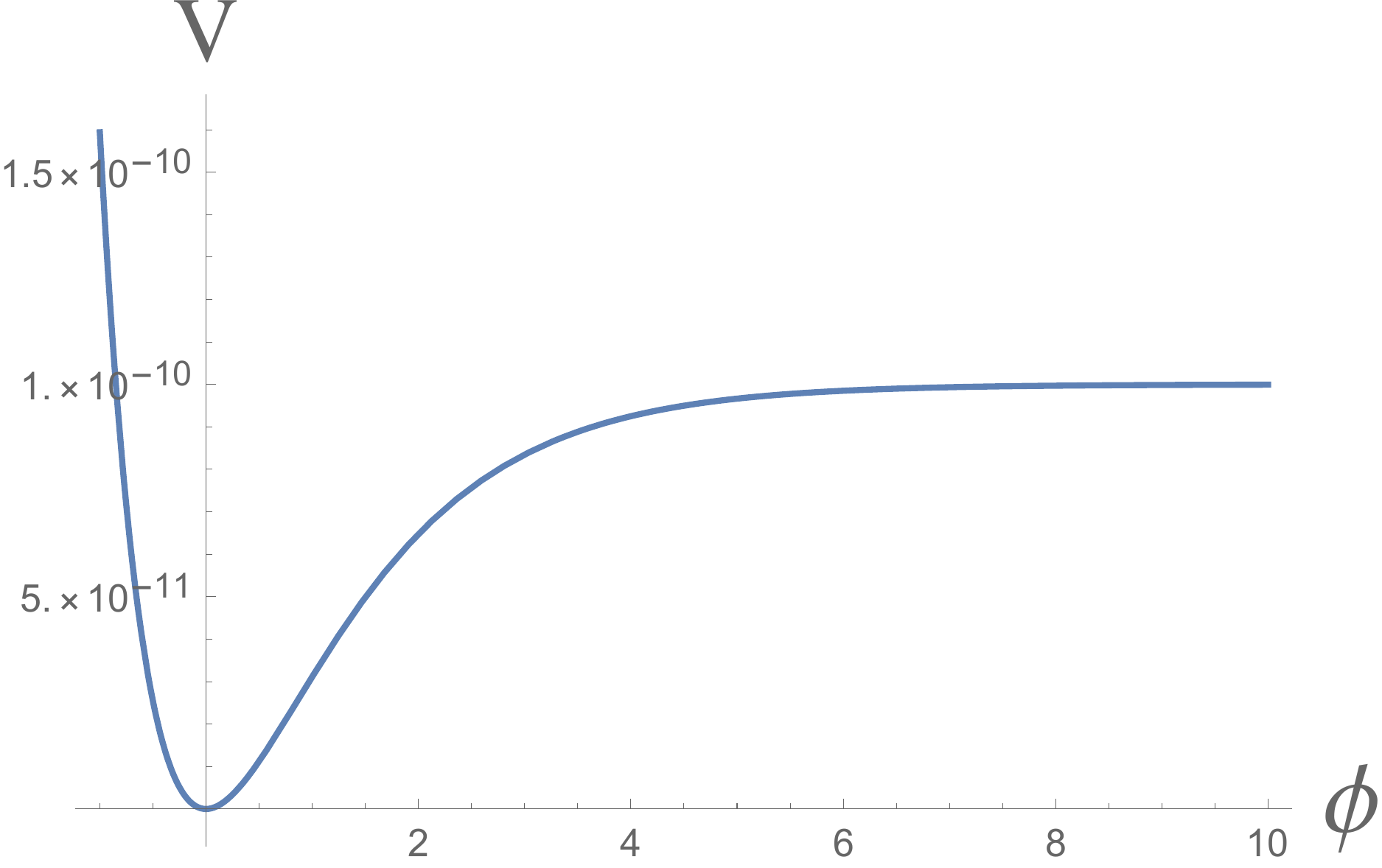}  \hskip 22pt
 \includegraphics[scale=0.39]{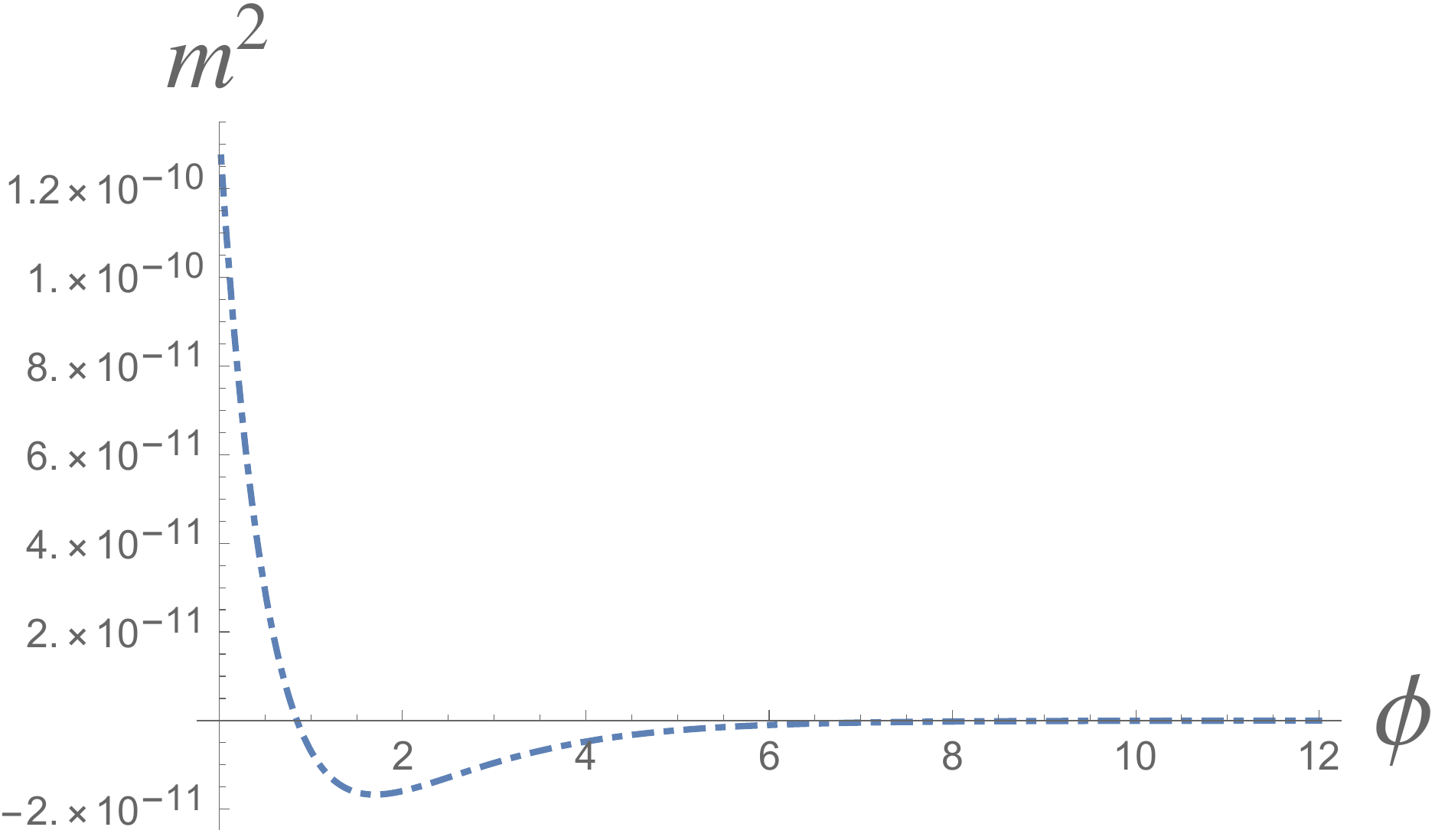}
\end{center}
\vspace{-.12cm}
\caption{\footnotesize The left panel shows the inflaton plateau potential of the $\alpha=1$ attractor model in equation \rf{3mod_Pot} for $\Lambda=0$ and $\mu =10^{-5}$. The right panel shows the inflaton mass squared. 
   }
\label{infl}
\end{figure}

The mass squared of the inflaton is  negative at the plateau of the potential and flips sign at $\phi\approx 0.8$, at which the absolute value of the inflaton mass squared shows a singular behavior in Fig.~\ref{infl}.

The heavy masses of the directions orthogonal to the inflationary trajectory ensure the validity of the sequestering mechanism, and we can safely use the effective description given by eq. \rf{3mod_Pot}. It is also worth emphasizing that the  effective K\"ahler potential acquires the different \K\, curvature radius, $\alpha_{1,2,3}=\frac13 \to \alpha_T=1$. Thus, the sequestering not only removes the heavy degrees of freedom, but also affects the inflationary  dynamics.

\section{Discussion}

In this paper we described and further developed a basic mechanism which allows us to obtain  inflationary models sequestered 
from physics at the Planckian energy scale. We illustrated this mechanism through several relatively simple examples. This mechanism was used in  \cite{Gunaydin:2020ric} to construct inflationary models in M-theory. 

At Step I, considering the case of one flat direction for simplicity, we find one massless supermultiplet and some number of heavy supermultiplets. At Step II we find that all heavy multiplets remain sufficiently heavy, with some split of masses inside the supermultiplet due to supersymmetry breaking. Meanwhile the massless multiplet of Step I is uplifted in a way which leads to a plateau potential for the real part of the superfield, i.e., for a single field inflation model. The axion field orthogonal to the inflationary trajectory is stabilized at vanishing value of the axion due to supersymmetry breaking and due to the choice of $V_{\rm infl}$ at Step II. Its mass is sufficiently heavy to decouple the axion from inflation despite these two fields being both massless at Step I.

In the follow-up investigation \cite{Kallosh:2021vcf}, we  will apply our methods to models where Step I is derived from IIB string theory. 
There are some specific features of such models. First of all, there are many scalar fields. We will study the STU model with 7 different moduli $S,\, T_1,\,T_2, \,T_3,\, U_1,\, U_2,\,U_3$. Secondly, there are many constraints on the structure of the superpotentials in such models. The general structure  of the superpotential $W_{\rm flux} (T^i)$ that we are using is described in \cite{Aldazabal:2006up}. It has terms depending on moduli starting from order zero up to order 5 in the fields. About 60 coefficients are arbitrary and are defined by various fluxes which are possible in 10D supergravity. These fluxes must satisfy about  100  tadpole cancellation conditions, which are basically Bianchi identities in the presence of local sources, such as D-branes and O-planes. It is not easy to find flux superpotentials which have supersymmetric Minkowski vacua and satisfy all tadpole conditions.

Fortunately, we have found such solutions in a class of superpotentials which are quadratic in the fields so that most of the tadpole cancellation conditions are satisfied trivially. We found solutions for the  remaining tadpole cancellation conditions. We have found vacua with one flat direction $T \equiv  S= T_1=T_2= T_3=U_1=U_2=U_3$ and with 3 flat directions $T_{(1)} \equiv U_1=U_3=T_1=T_3$,  $ T_{(2)} \equiv S=U_2$ , $T_{(3)} \equiv T_2$.  

Then, using the methods outlined in the present paper, we uplifted these flat directions and studied inflation in these models. Just as in the models discussed in  the present paper, the high energy scale parameters appearing in string theory and the heavy masses do not affect the inflationary dynamics. The main consequence of string theory inherited by the inflationary models is the hyperbolic geometry of the moduli space, which helps to develop $\alpha$-attractor inflationary models with the discrete set of possible values  $3\alpha=7,6,5,4,3,2,1$.

\section*{Acknowledgement}
We are grateful to  G. Dall'Agata, S. Ferrara,  R. Flauger,  D. Roest and   A. Van Proeyen   for stimulating discussions.  RK and AL are supported by SITP and by the US National Science Foundation Grant  PHY-2014215, and by the Simons Foundation Origins of the Universe program (Modern Inflationary Cosmology collaboration).  YY is  supported by JSPS KAKENHI, Grant-in-Aid for JSPS Fellows JP19J00494. TW is supported in part by the US National Science Foundation Grant  PHY-2013988.

\bibliographystyle{JHEP}
\bibliography{lindekalloshrefs}
\end{document}